# Optically Probing Unconventional Superconductivity in Atomically Thin Bi$_2$Sr$_2$Ca$_{0.92}$Y$_{0.08}$Cu$_2$O$_{8+\delta}$

*Yunhuan Xiao[1,2,#], Jingda Wu[1,2,#], Jerry I Dadap[1,2], Kashif Masud Awan[2], Dongyang Yang[1,2], Jing Liang[1,2], Kenji Watanabe[3], Takashi Taniguchi[4], Marta Zonno[1,2], Martin Bluschke[1,2], Hiroshi Eisaki[5], Martin Greven[6], Andrea Damascelli[1,2], Ziliang Ye\*[1,2]*

[1] Department of Physics & Astronomy, University of British Columbia, Vancouver, British Columbia V6T 1Z1, Canada

[2] Quantum Matter Institute, University of British Columbia, Vancouver, British Columbia V6T 1Z4, Canada

[3] Research Center for Functional Materials, National Institute for Materials Science, 1-1 Namiki, Tsukuba 305-0044, Japan

[4] International Center for Materials Nanoarchitectonics, National Institute for Materials Science, 1-1 Namiki, Tsukuba 305-0044, Japan

[5] Nanoelectronics Research Institute, National Institute of Advanced Industrial Science and Technology, Tsukuba, Ibaraki 305-8568, Japan

[6] School of Physics and Astronomy, University of Minnesota, Minneapolis, MN 55455, USA

**ABSTRACT**

Atomically thin cuprates exhibiting a superconducting phase transition temperature similar to bulk have recently been realized, although the device fabrication remains a challenge and limits the potential for many novel studies and applications. Here we use an optical pump-probe approach to noninvasively study the unconventional superconductivity in atomically thin $Bi_2Sr_2Ca_{0.92}Y_{0.08}Cu_2O_{8+\delta}$ (Y-Bi2212). Apart from finding an optical response due to the superconducting phase transition that is similar to bulk Y-Bi2212, we observe that the sign and amplitude of the pump-probe signal in the atomically thin flake vary significantly in different dielectric environments depending on the nature of the optical excitation. By exploiting the spatial resolution of the optical probe, we uncover the exceptional sensitivity of monolayer Y-Bi2212 to the environment. Our results provide the first optical evidence for the intralayer nature of the superconducting condensate in Bi2212, and highlight the role of double-sided encapsulation in preserving superconductivity in atomically thin cuprates.

The successful isolation of monolayer (ML) Bi$_2$Sr$_2$CaCu$_2$O$_{8+\delta}$ (Bi2212) has opened up many exciting possibilities for studying two-dimensional (2D) superconductivity and fabricating novel devices using atomically thin high-$T_c$ superconductors (2D-HTSCs)[1]. Recent experiments have revealed that novel phenomena such as the interfacial Josephson effect and superconducting diode effect can emerge in artificially twisted Bi2212 stacks[2–5]. Theoretically, it has been predicted that such twisted stacks can host more exotic physics, including topological superconductivity with broken time-reversal symmetry[6–8]. However, fabricating high-quality cuprate devices and probing their intrinsic properties through transport measurements remains a challenging task. One of the primary challenges lies in the susceptibility of atomically thin Bi2212 to degradation during conventional nanofabrication processes due to oxygen dopant loss and reaction with moisture[1,9–11]. To mitigate these issues, stencil mask lithography and limited-heating evaporation have been employed to reduce the damage of electrical contact fabrication in relatively thick flakes[10,12]. Additionally, a cold welding approach has been developed to preserve the ML sample quality[1], although this method requires implementing a series of intricate processes in a stringent environment. In this study, we propose an alternative approach to investigate the superconductivity in atomically thin Bi2212 using ultrafast optical pump-probe spectroscopy, a noninvasive probe with micron-sized spatial resolution that allows us to study the interaction of 2D-HTSCs with different local environments.

Ultrafast optical pump-probe spectroscopy has been successfully applied to study superconductivity in bulk HTSCs[13]. In these measurements, an ultrafast pump pulse is firstly applied to samples of over 100-nm thickness to break the Cooper pairs. The transiently formed non-equilibrium quasiparticles alter the optical conductivity and reflectivity, which is measured with an ultrashort probe pulse with a variable time delay following the pump pulse. Remarkably, the superconducting phase transition can be observed by monitoring the decay time constant $\tau$ of the transient reflection signal. In the superconducting phase, $\tau$ is several times longer than at temperatures above $T_c$. This contrast has been understood as a phonon bottleneck effect in superconductors[14,15], where photoexcited quasiparticles quickly relax to the superconducting gap edge and recombine while emitting numerous gap-frequency phonons (GFPs) (Figure 1a). These GFPs can further break other Cooper pairs, re-excite the quasiparticles, and generate new GFPs, until the energy dissipates through anharmonic decay[13,16]. As a result, $\tau$ is usually a few picoseconds long in the superconducting phase and is often associated with the GFP decay time. In contrast, such a bottleneck effect is absent in the non-superconducting phase, where $\tau$ becomes shorter than 1 ps[17,18]. Although the phonon bottleneck equations are derived for the isotropic *s*-wave superconducting gap, the photoexcitation and relaxation of quasiparticles in *d*-wave cuprates are expected to take place mainly at antinodes[16,19–21], where the gap size is maximized and used in the phonon bottleneck effect.

In this work, we push the limit of this optical sensing technique and apply it to atomically thin Y-Bi2212. We find that the majority of the bulk behaviors persist in the atomically thin limit. Specifically, we find a similar change of the decay time constant $\tau$ near the phase transition and the ability of a strong pump pulse to induce a non-superconducting phase below $T_c$. These agreements with the bulk observations elucidate the 2D nature of Cooper pairs in Bi2212[22–28], where the superconductivity is predominantly hosted by two coupled $CuO_2$ planes. On the other hand, we find atomically thin Y-Bi2212 to be easily affected by the local environment. Enabled by the spatial resolution offered by our optical technique, we are able to resolve the transient reflection signal in different local environments, which allows us to resolve the complex permittivity contrast in the superconducting and non-superconducting components of the pump-probe signal. In addition, we find that a suspended Y-Bi2212 monolayer can be extremely sensitive to the residual gas through either the upper or lower exposed surface, and a double-sided encapsulation proves effective in preserving the superconductivity.

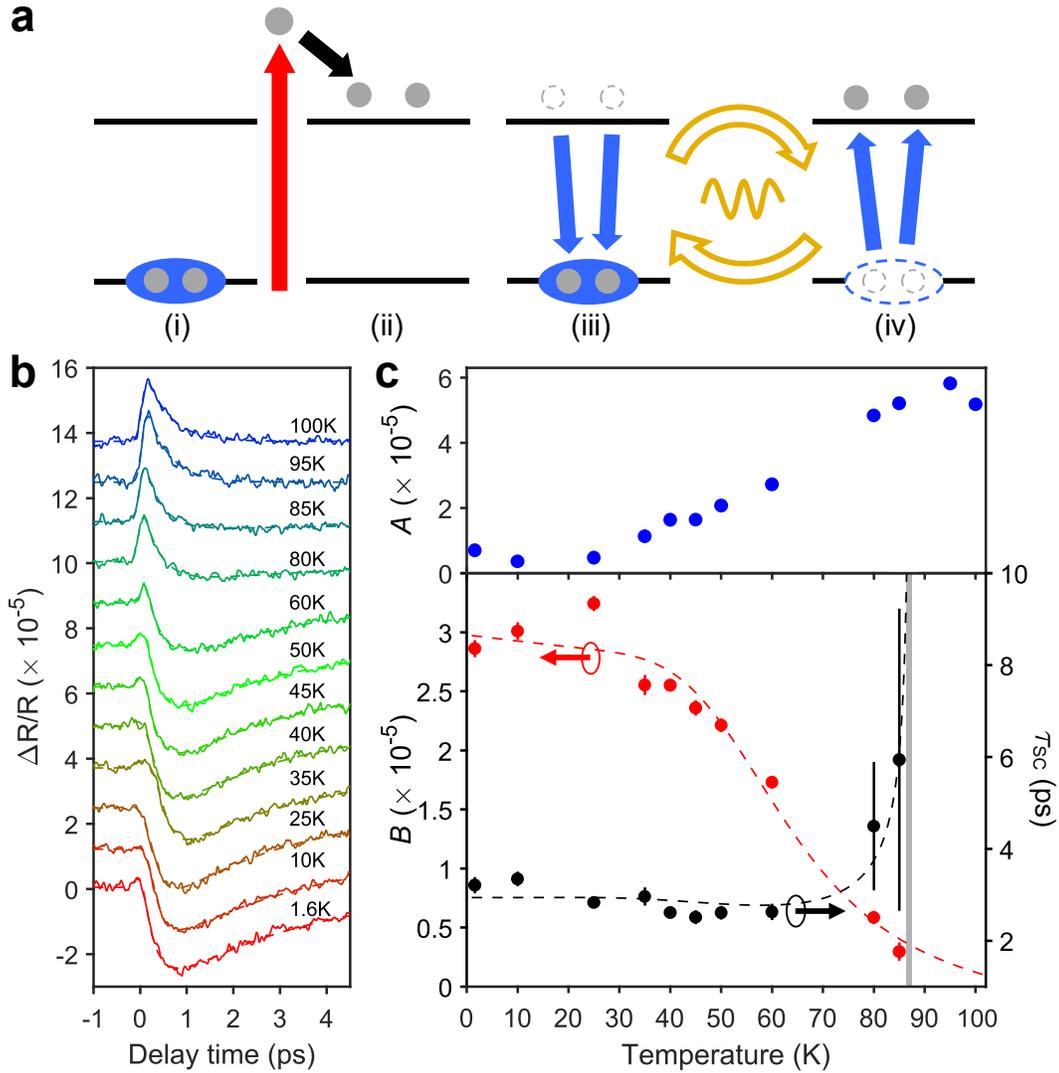

**Figure 1.** (a) The phonon bottleneck effect involved in the relaxation of optical excitations in high-$T_c$ superconductors. (i) Photoexcitation (red arrow) breaks Cooper pairs and generates quasiparticles above the superconducting gap. (ii) Thermalization occurs as quasiparticles relax to the states near the edges of the superconducting gap. (iii) Quasiparticles can recombine into Cooper pairs only by emitting phonons with the gap energy. (iv) The gap frequency phonon

(GFP) breaks another Cooper pair and brings the system back to the condition of (iii). The cyclical process between (iii) and (iv) continues until GFPs decay through the anharmonic processes or diffuse out of the probing spot. (b) Temperature dependence of the pump-probe response of a four-layer (4L) Y-Bi2212 flake measured at a pump fluence of 44.5 µJ/cm$^2$. Curves for $T > 1.6$ K are shifted vertically. Dashed lines are the fit results based on a single exponential (for 1.6 K, 95 K and 100 K) or a bi-exponential function (for all other temperatures). (c) Top panel: extracted PG component amplitude $A$ (blue circles) at each temperature. The PG component emerges near 40K and becomes larger with increasing temperature. Bottom panel: extracted SC component amplitude $B$ (red circles) and decay time constant $\tau_{SC}$ (black circles) at each temperature. The uncertainty in $\tau_{SC}$ becomes very large near $T_c$ as the SC component vanishes near the phase transition. Fitted curves are shown by dashed lines. The gray vertical bar highlights the $T_c$ from $\tau_{SC}$ fitting from the signals of the 4L sample.

We start by measuring the temperature dependence of the pump-probe response in a four-layer (4L) sample. Following the convention in the field, each monolayer is defined as half of the Bi2212 unit cell and contains two CuO$_2$ planes separated by a Ca plane[1,29] (Figure S1). We use ~100-fs pump (1200 nm) and probe (800 nm) pulses for our optical pump-probe experiments. Details about the experimental methods can be found in the Supporting Information. We first identify the characteristics of the time-resolved signals at temperatures both below and above the bulk crystal $T_c$ with a pump fluence of 44.5 µJ/cm$^2$. At base temperature (1.6 K), the pump reduces the transient reflection of the probe. The negative signal initially rises in ~100 fs and then gradually recovers over a few ps. At 100 K, the pump-probe signal is positive and the decay occurs mostly within one ps (Figure 1b). To quantify this contrast, a single exponential decay convoluted with a Gaussian pump pulse shape function is used to extract the respective time constants. The time constant $\tau$ of the slow recovery at 1.6 K is about 3.2 ps, eight times slower than the fast decay at 100 K (~ 0.4 ps). These observations agree with the so-called superconducting (SC) and pseudogap (PG) components in bulk HTSCs, where the values of the time constant are $\tau_{SC} \approx 2.5$ ps and $\tau_{PG} \approx 0.5$ ps respectively[17], suggesting that superconductivity is preserved and the phonon bottleneck effect persists in atomically thin Bi2212. We further confirm the strong correlation between the slow-decaying pump-probe signal and superconductivity by carrying out optical and electrical four-probe resistance measurements in the same 10-layer sample. Detailed information is available in the Supporting Information.

The results at intermediate temperatures are more complicated as they involve both positive and negative components due to the pump-induced phase transition of some regions from the SC phase to the PG phase. We fit these signals with the combination of two exponential decays[30], $Ae^{-t/\tau_{PG}} - Be^{-t/\tau_{SC}}$, where $A$ ($B$) and $\tau_{PG}$ ($\tau_{SC}$) are the amplitude and time constant of the PG (SC) component, respectively, and the result of this analysis is shown in Figure 1c. As temperature increases, the amplitude of the SC component diminishes and falls below the detection limit at 95 K and 100 K. Assuming that $B$ is proportional to the photoexcited Cooper pair density, we can fit $B(T)$ with a two-temperature model previously developed for bulk HTSCs[16]. On the other hand, $\tau_{SC}$ increases with increasing temperature near $T_c$, which can be understood in terms of the Ginzburg-Landau theory of second-order phase transitions[31]. As the temperature approaches $T_c$, the restoring force corresponding to the derivative of the system's free energy with respect to the SC order parameter diminishes, and therefore the relaxation to equilibrium after a sudden pump becomes infinitely long. The seeming divergent behavior of $\tau_{SC}$ agrees with previous pump-probe studies of bulk HTSCs[16,17,30,32]. By fitting the $\tau_{SC}$ with the phonon decay lifetime in the superconducting state[16], we acquire a $T_c$ of ~ 87 K. The bulk $T_c$ of ~ 92 K (slightly underdoped Y-Bi2212) falls within the uncertainty of our measured $T_c$ (see Supporting Information for details). The similar overall temperature dependence and superconducting transition temperature confirm that the SC gap in Bi2212 is not affected by the reduced dimension.

The existence of both SC and PG components at intermediate temperatures can be explained by a photoinduced phase transition (PIPT) triggered by a pump pulse above a certain threshold fluence ($\Phi_{th}$)[18,30,33–36]. At low pump fluence, only a small number of Cooper pairs become quasiparticles and their relaxation is governed by the phonon bottleneck effect. When Cooper pairs are pumped strongly with a fluence exceeding the threshold, in our case $\Phi_{th} \approx 50$ μJ/cm$^2$ at 1.6 K, some regions of the material go through a phase transition and enter the pseudogap phase[34]. Such a transient phase was previously attributed to a collection of Cooper pairs without phase coherence[25] and argued to be non-thermal[18,30]. An alternative explanation for the coexistence of both components is a percolation model, which considers a broad distribution of gaps and nanoscale superconducting patches that proliferate in the material upon cooling[37,38], and increased fluence may convert the patches into non-superconducting ones. This description was introduced also in relation to the overall superconducting phenomenology of cuprates and oxide superconductors[39].

The pump-induced mixed phase can be clearly seen in our 4L sample even at 1.6 K (Figure 2a). As the pump fluence increases, the dynamics gradually changes from a single-exponential recovery to a double-exponential decay at intermediate temperatures. When the fluence is small, the fitted amplitude $B$ increases linearly with the fluence and the amplitude $A$ is negligible (Figure 2b). Above $\Phi_{th} \approx 50$ μJ/cm$^2$, the strong pump saturates the SC component while the PG component appears and dominates the signal in the first few hundreds of femtoseconds. Our reported temperature dependence in Figure 1b is measured with a pump fluence of 44.5 μJ/cm$^2$, as indicated by the vertical dashed line, which is close to $\Phi_{th}$. Nevertheless, the SC gap decreases as the temperature increases, which reduces $\Phi_{th}$. Consequently, the intermediate temperature dependence results reported in Figure 1b are likely in the early saturation regime, which leads to the observed PG component. This saturation may also slightly distort the fitted amplitude of the SC component, but the time constant and the retrieved $T_c$ should not be influenced, because the decay time constant may only be affected by the fluence at the lowest temperature due to a possible two-particle kinetics[15,18,40], while the saturation occurs at higher temperatures.

Interestingly, the observed low-temperature saturation threshold (~50 μJ/cm$^2$) is comparable to that in bulk Bi2212, where the collapse of superconductivity has been reported for pump fluences ranging from 14 μJ/cm$^2$ to 70 μJ/cm$^2$[13,30,33–36]. Given the dielectric functions of Y-Bi2212[41], SiO$_2$ (285 nm), and hBN (15 nm), we calculate and find that the average absorption per layer at 1200 nm (pump wavelength) in both bulk and the 4L Y-Bi2212 are similar. Therefore, we conclude that the intrinsic PIPT threshold fluence is not changed in the atomically thin Bi2212, implying that the phase coherence of the SC condensate is not related to the coupling between layers.

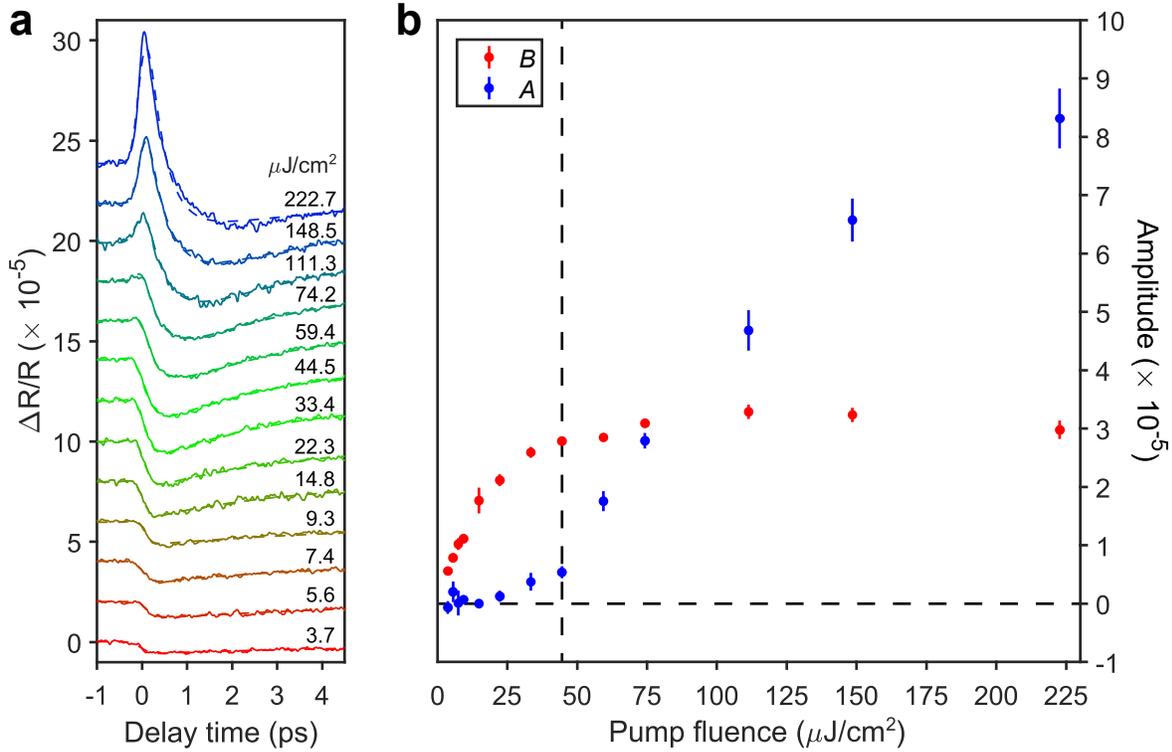

**Figure 2.** (a) Fluence dependence of the pump-probe response of the 4L Y-Bi2212 sample at 1.6 K. (b) Extracted PG and SC component amplitudes, labeled by *A* (blue circles) and *B* (red circles) respectively. The vertical dashed line indicates the fluence (44.5 μJ/cm$^2$) applied for the temperature dependence measurement (Figure 1b).

So far, we have demonstrated that the atomically thin Bi2212 manifests many similar optical properties as the bulk. In addition to the above findings, we have discovered novel properties that arise from the atomically thin nature of 2D-HTSC. Specifically, we find that the pump-probe response of a thin flake can be significantly modified by the environment. This environmental contrast provides us with the ability to resolve the complex permittivity change associated with different optical excitations.

To this end, we conduct experiments on flakes prepared on substrates prepatterned with hole arrays. The SiO$_2$ substrate and holes provide large permittivity contrast at the optical frequency, which allows us to measure different pump-probe responses at 1.6 K. To avoid complications due to the hBN encapsulation layer, we use a relatively thick Y-Bi2212 flake (40 nm), which is more robust to oxygen dopant loss, but still thinner than the optical skin depth[42] (~ 100 nm). In the area supported by the substrate, we observe a response similar to that of the 4L flake: with a strong pump above the threshold fluence, the signal shows a slow SC component with a negative sign and a fast PG component with a positive sign (top panel in Figure 3a). Interestingly, in the suspended area, the slow SC component turns positive, and its amplitude becomes more than ten times stronger, while the fast PG component remains positive with an about three times stronger amplitude (bottom panel in Figure 3a). Since the two probe spots are only a few micrometers apart, this sign change is unlikely due to the local oxygen doping level difference between these two sample areas, in contrast to the case for bulk samples where the doping level usually determines the sign change behavior[41,43].

Through a series of calculations based on the transfer-matrix formalism, we conclude that the contrast stems from the nature of the optical excitations. We simulate the sign and amplitude of the pump-probe response using the transfer-matrix method based on each layer's dielectric function and dimension. The pump-induced optical conductivity change of the sample is modeled as a relatively small permittivity change ($\leq 10^{-4}$), either in the real ($\varepsilon_1$) or imaginary ($\varepsilon_2$) part of the sample layer, while other layers are unaffected by the pump. To compare with the experiment, we normalize the reflectivity change of the probe beam between the pump on and off conditions to the absolute reflectivity. As the calculation result in Figure 3b shows, the sign and amplitude of $\Delta R / R$ are significantly affected by an interference effect. In all scenarios, $\Delta R /$ $R$ oscillates with the film thickens, with a periodicity matching the half-$\lambda$ condition for the probe light in Bi2212. For a flake of a certain thickness, the sign of $\Delta R / R$ is dependent on the nature of the permittivity change and local dielectric environment. In the thick limit (> 1μm), the probe beam can barely penetrate through the Bi2212 layer, resulting in a vanishing interference effect and a convergence to the bulk limit.

The calculated $\Delta R / R$ contrast between the supported and suspended area is summarized in a ratio plot (Figure 3c). When the pump-induced complex permittivity change is dominated by the imaginary part ($\Delta\varepsilon_1 / \varepsilon_1$ close to zero), $\Delta R / R$ exhibits the same sign in both areas, mirroring the behavior of the fast PG component. Under other conditions, $\Delta R / R$ changes its sign, similar to the behavior of the slow SC component. These correspondences are supported by a phenomenological model developed to describe the dielectric response of Y-Bi2212[41]. According to this model, the PG component is attributed to a pump-induced broadening of the Drude peak so that mainly the $\varepsilon_2$ contribution is transiently modified. On the other hand, the SC component stems from the pump-induced changes in the Lorentzian oscillator of the interband transition at 1.5 eV, leading to changes in both $\varepsilon_1$ and $\varepsilon_2$. Our probe light measures a different combination of $\varepsilon_1$ and $\varepsilon_2$ in the supported area and the suspended area. As a result, the PG component (with mainly changes in $\varepsilon_2$) is associated with enhancement in the transient reflectivity in both areas, while the SC component (with changes in both $\varepsilon_1$ and $\varepsilon_2$) induces opposite $\Delta R / R$ changes between the two areas. The unique contrast in the permittivity change between the PG and the SC components is interpreted as an indication of an unconventional superconductivity-induced carrier kinetic energy loss[41].

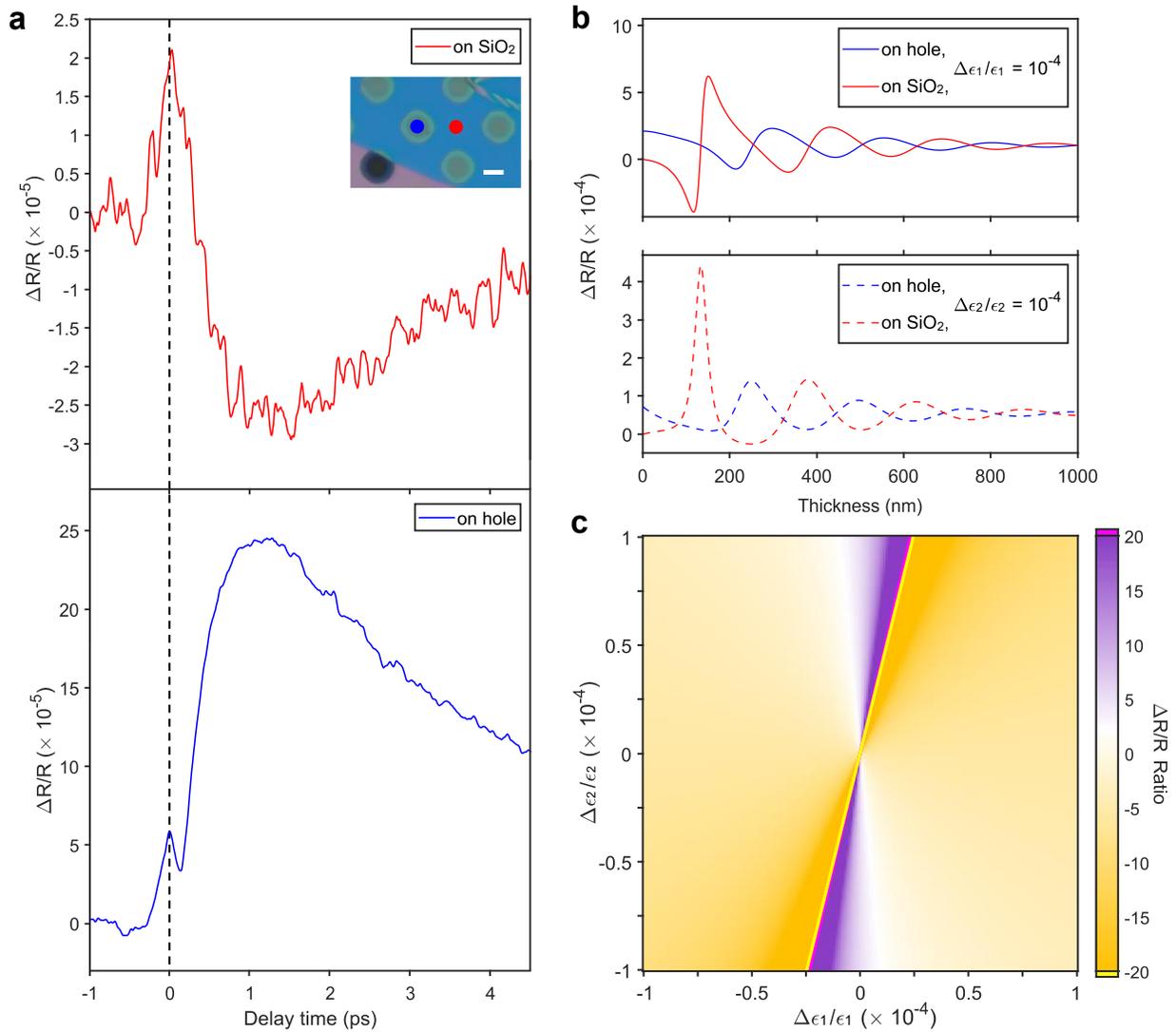

**Figure 3.** (a) The pump-probe response of a suspended Y-Bi2212 sample showing a large contrast compared to the area supported by SiO$_2$. The measured spots in the suspended and supported areas are indicated by blue and red dots in the optical microscope image (inset), respectively. The scale bar is 5 μm. The supported area shows a positive fast and a negative slow component, while both the fast and slow responses in the suspended area are positive. The vertical dashed line is the zero time delay. (b) The calculated flake thickness dependence of normalized transient reflectivity from the suspended (blue) and supported (red) areas. The top panel shows the contrast when the permittivity change is real, and the bottom panel corresponds

to a purely imaginary permittivity change. (c) Calculated ratio of transient reflectivity of the suspended and supported areas for a 40-nm thick Y-Bi2212 flake. The purple and gold areas represent the sign-conserving and sign-reversing conditions, respectively.

In the second example, we push the thickness limit and apply the pump-probe technique to a monolayer (ML) Y-Bi2212 flake prepared on a prepatterned substrate (Figure 4a). In Figure 4b, the area left to the pink dashed line contains a ML, half of which is covered by hBN, which appears blue in the optical microscope image. The pump-probe responses of three representative spots are measured at 1.6 K with a μm-level spatial resolution, as illustrated in Figure 4c. Clearly, only the ML sandwiched between hBN and $SiO_2$ (red dot) remains superconducting, showing a slow-decaying dynamics similar as the SC component in thicker flakes. A fluence of 74 μJ/cm$^2$ is utilized here to achieve a sufficient signal-to-noise ratio on the ML, thus both SC and PG components are observed. The signs of both components are positive, potentially due to the local dielectric environment of the sample. The SC component persists up to ~ 60 K, which we identify as the $T_c$ of the ML sample.

In contrast to the red-dotted position, the sample at the other two locations shows no response above the noise floor at base temperature, indicating a complete quench of the superconductivity, which cannot result from the dielectric environment difference according to our transfer matrix calculations. We expect the null signals to have the following two potential causes: (i) The sample becomes amorphous after reacting with the residual water vapor in a gaseous environment[10]; (ii) The sample becomes insulating as the oxygen dopant leaves over time at room temperature[1,11]. Both conditions are equally possible for the exposed area (green dot), because the sample without hBN coverage is directly in contact with the ambient environment before being loaded into the cryostat. For the hBN-covered suspended area (black dot), the oxygen dopant loss is more likely a reason for the loss of superconductivity since the $N_2$ trapped in the hole should contain little moisture or oxygen from the glovebox. Both results indicate that monolayer Y-Bi2212 is extremely sensitive to the environment, and it is crucial to encapsulate them from both sides. In our experiment, hBN and $SiO_2$ provide effective barriers for hampering the migration of water vapor and oxygen dopants. A similar protection has also been demonstrated using graphene[9].

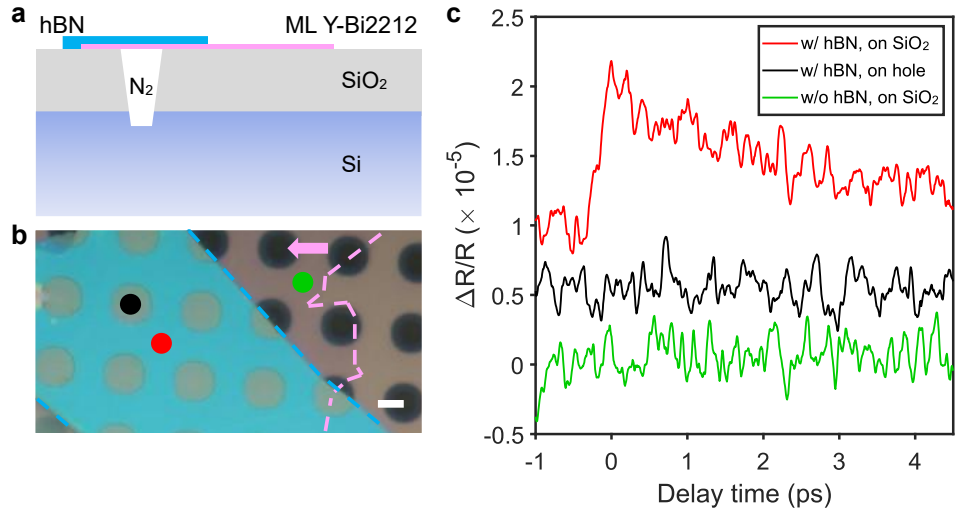

**Figure 4.** (a) Schematic of a ML Y-Bi2212 sample prepared on a prepatterned substrate. A piece of hBN (blue) covers the ML Y-Bi2212 sample (pink) as an overlayer. (b) Corresponding optical microscope image. Dashed lines outline the boundaries for hBN (blue) and ML Y-Bi2212 (left to the pink dashed line, as indicated by the arrow). The three probing spots are marked by black (suspended ML covered by hBN), red ($SiO_2$ supported area covered by hBN), and green (uncovered supported area) dots. The scale bar is 5 μm. (c) Pump-probe responses at 1.6 K for the ML sample, corresponding to the three dots in (b). Superconductivity is only preserved in the hBN covered supported area.

To summarize, we have studied the superconducting phase transition in high-quality Y-Bi2212 samples of atomic thickness using an optical pump-probe technique. The optical approach provides a noninvasive way to probe the superconductivity with a micrometer spatial resolution and to identify the intrinsic $T_c$, which is close to the bulk limit. We find that the decay time constant shows a divergent-like behavior near $T_c$, and observe a pump-induced phase transition at high fluence, consistent with previous reports for bulk samples. These findings indicate that superconductivity persists in Bi2212 with reduced dimension, confirming the 2D nature of the electronic correlations of the superconducting condensate. Furthermore, we uncover significant effects of the local environment and flake thickness on the pump-probe response of atomically thin samples. We spatially resolve the pump-probe signals of different signs and amplitudes, which can be attributed to different optical excitations in different local dielectric environments. Remarkably, we find that the superconductivity in the monolayer sample is extraordinarily sensitive to exposed surfaces, which highlights the importance of protective encapsulation layers. Our understanding of optical properties of atomically thin cuprates may pave the way for their potential applications in optoelectronics and photon-based quantum computing[44,45].

**ASSOCIATED CONTENT**

**Supporting information**

Supporting Information: Detailed experimental methods, temperature dependence analysis, and the additional transport experiment (PDF)

## AUTHOR INFORMATION

**Corresponding author**

*Ziliang Ye, E-mail: zlye@phas.ubc.ca

**Author contributions**

# These authors contribute equally to this paper. Z.Y. conceived the idea and managed the project. Y.X. fabricated the samples with the assistance from D.Y., J.L., and K.A.. Y.X., J.W. and J.D. performed the measurements. M.Z., M.B., A.D., H.E., M.G., K.W., and T.T. provided the bulk crystals. Y.X. and Z.Y. performed the analysis and wrote the manuscript with input from all other authors.

**Notes**

The authors declare that they have no conflict of interest.

## ACKNOWLEDGEMENTS

We acknowledge the fruitful discussions with Frank Zhao and Hsiang-Hsi Kung, as well as the technical support from Giorgio Levy and Vinit Doke. This research was undertaken thanks in part to funding from the Max Planck-UBC-UTokyo Center for Quantum Materials and the Canada First Research Excellence Fund, Quantum Materials and Future Technologies Program and Gordon and Betty Moore Foundation's EPiQS Initiative (GBMF11071). The work at the University of Minnesota was funded by the U.S. Department of Energy through the University of Minnesota Center for Quantum Materials, under Grant No. DE-SC0016371. This project was also funded by the Natural Sciences and Engineering Research Council of Canada (NSERC); the Canada Foundation for Innovation (CFI); the Department of National Defence (DND); the British Columbia Knowledge Development Fund (BCKDF); the Canada Research Chairs Program (Z.Y., A.D.); and the CIFAR Quantum Materials Program (A.D.).